\documentclass[12pt,hyper,notoc]{JHEP3}
\pdfoutput=1

\usepackage{amssymb,amsthm,amscd,calc}
\usepackage{graphicx}
\usepackage{cite}
\usepackage{subfloat}
\usepackage{epsfig}

\newcommand\be{\begin{equation}}
\newcommand\ee{\end{equation}}
\newcommand{\bea}{\begin{eqnarray}}
\newcommand{\bean}{\begin{eqnarray*}}
\newcommand{\eean}{\end{eqnarray*}}
\newcommand{\eea}{\end{eqnarray}}

\newcommand{\bead}{\begin{align}}
\newcommand{\eead}{\end{align}}

\title{AdS Phase Transitions at finite $\kappa$}

\author{ Gianni Tallarita \\ \\ 
Queen Mary University of London \\
Centre for Research in String Theory \\
Department of Physics, \\
Mile End Road, London, E1 4NS, UK. \\ 
\\
Email: \email{ G.Tallarita@qmul.ac.uk}}

\abstract{We investigate the effect of adding a Chern-Simons term coupled to an axion field to $SU(2)$ Einstein-Yang-Mills in a fixed $AdS_4$/Schwarzschild background. We show that, when the axion has no potential, there is a phase transition between a Reissner-Nordstrom black-hole and one with a non-abelian condensate as per the vanishing Chern-Simons case. Furthermore, by giving the axion field a mass, one observes a phase transition between a Reissner-Nordstrom black-hole with axion hair to a ``superconducting" phase which also has a non-trivial axion profile.  We are able to perform a preliminary analysis for this interesting case and observe that we can shift the critical temperature at which the phase transition occurs and observe interesting features of the order parameter scaling form.}

\preprint{QMUL-PH-10-19}
\begin{document}

\section{Introduction}

\indent	In recent years the subject of phase transitions in black holes has been extensively studied. Much is to be attributed to the development of $AdS/CFT$ \cite{Maldacena} within the context of condensed matter systems, especially in holographic superconductivity \cite{hartnoll1}, \cite{hartnoll2} (see \cite{herzog2} \cite{hartnoll3} \cite{Horowitz} for comprehensive reviews on the subject). In such models superconductivity is described by a phase transition of an asymptotically $AdS$ black hole which admits both a Reissner-Nordstrom (RN) and an $AdS$/Schwarschild solution. In the initial proposal an abelian condensate is described by a charged scalar field acquiring a VEV on the boundary of the $AdS$ space or, in the dual gravitational picture, the black-hole developing scalar hair. The model was extended to include phase-transitions of $AdS$ black-holes involving non-abelian condensates \cite{gubser1} and these led to phenomenologically promising models of p-wave superconductivity \cite{gubser2}. Since then, models of p-wave holographic superconductivity have been widely studied \cite{russo} \cite{zeng} \cite{Cai} \cite{gubser3} \cite{zeng2} \cite{erdmenger1} \cite{erdmenger2}. \\
\indent	In \cite{tallarita} it was shown that one can also observe characteristics of Chern-Simons (CS) interactions in superconductivity described by abelian condensates by coupling a CS term to a neutral axion field in the four-dimensional Einstein-Maxwell action. In this case the condensate is still described by an external scalar field which doesn't couple to the axion field, hence the condensate profiles are identical to those with a vanishing CS term. Vortex solutions of the system lead to properties of pure CS systems such as the magnetic field peaking outside of the core of the vortex. In this paper we wish to make progress towards including CS effects in four dimensional $AdS$ black-hole phase transitions involving non-abelian condensates. Our mechanism will be very similar to the previously mentioned case: we will couple an $SU(2)$ CS term to an axion field, solve the system in the bulk and project it to the boundary where we hope to observe interesting features of the dual field theory. In this case however, since the condensate is provided by the gauge field itself, we are effectively coupling the axion field directly to the condensate through the CS term and this will lead to interesting novel observations on the profile of the order parameter and the space of possible solutions describing the phase transition. Unlike the case of the abelian condensate we don't expect to observe CS characteristics in the boundary field theory itself, our approach here is to investigate the holographic effects on this theory observed via the CS coupling in the bulk. The hope is that through this study one could in the future investigate the full effects of the CS term on a model of p-wave superconductivity.\\
\indent	The paper is organised as follows: in section 2 we introduce the system consisting of $SU(2)$ Einstein-Yang-Mills with the inclusion of an axion field $\theta$ coupled to a CS term. We make an ansatz for the gauge field, derive the equations of motion of the system and by expanding the fields at the $AdS$ boundary discuss the relevant dual thermodynamical variables of interest to the problem. Section 3 is devoted to numerical solutions to the above system. There are two parts: the first in which the axion potential is set to zero and the second in which we switch on a mass term for the axion field. We show here that these lead to very different results. In section 4 we calculate and evaluate the free energy of the different phases by using the $AdS/CFT$ dictionary and in section 5 we provide short analytical results of the $T\approxeq T_c$ region which give useful general behaviours of quantities of interest. Finally, in section 6, we provide a brief summary of the results obtained and point the reader to directions for future work. 
 
\section{The System}

Our starting point will be $SU(2)$ Einstein-Yang-Mills with a Chern-Simons term coupled to a dimensionless axionic field $\theta$ in the $AdS_4/Schwarzschild$ ($AdS/Sch$) background.

\bea
S =\frac{1}{2k_4^2} \int dx^4\sqrt{-G}\left(R-\Lambda+\frac{1}{g^2}\mathcal{L}\right)
\eea

where
\be
\mathcal{L}=-\frac{1}{4}Tr\left(F_{\mu\nu}F^{\mu\nu}\right)+\frac{\kappa}{\sqrt{-G}}\theta Tr\left(F\wedge F\right)+\lambda^2\partial_\mu\theta\partial^\mu\theta+V(\theta)
\ee

where $F^a_{\mu\nu}=\partial_\mu A_\nu^a-\partial_\nu A_\mu^a+\epsilon^{abc}A_\mu^bA_\nu^c$,  the cosmological constant $\Lambda=-\frac{6}{L^2}$, $L$ is the AdS radius, $k_4^2=8\pi G_N$, $\lambda$ is a dimension one constant (usually called the axion decay constant) which we set to 1 manually and $V(\theta)$ is the axion potential which for the moment we will leave unspecified. $\kappa$ is a constant useful in keeping track of the relative Chern-Simons contribution to the action. It is important here to note that this set-up manifestly violates parity symmetry under the transformation $\theta \rightarrow -\theta$. \newline

We work in the non-backreacting limit of large $g$ where we can take the $AdS/Sch$ ansatz for the metric

\be
ds^2=\frac{r^2}{L^2}\left[-\left(1-\frac{r_h^3}{r^3}\right)dt^2+dx^2+dy^2\right]+\frac{L^2}{r^2}\frac{dr^2}{1-\frac{r_h^3}{r^3}},
\ee

in which $r_h$ indicates the position of the black hole horizon. We take the gauge field $A=A_\mu^a\tau^a dx^\mu$ where $\tau^a$ are the generators of the $SU(2)$ algebra such that $[\tau^a,\tau^b]=\epsilon^{abc}\tau^c$. The effective boundary Chern-Simons coupling term is $\kappa\theta$ evaluated on the boundary and we must impose that this is quantized to work with a reasonable Chern-Simons theory (see \cite{gerald}), this will put restrictions on the values of $\theta$ at the horizon as will be shown later.\\

In \cite{gubser1} it was shown that, for the case of $\kappa=0$ a $p+ip$ gauge field ansatz
\be\label{ansatz}
A=\phi\tau^3dt+\omega\left(\tau^1dx+\tau^2dy\right)
\ee

leads to a second order phase transition from a RN anti-de Sitter black hole to one with non-abelian condensates, with non zero $\omega$ acting as the condensate. However in \cite{gubser2} the authors argue that these backgrounds are unstable to small perturbations that turn them into the less isotropic $p$-wave backgrounds, at least close to $T=T_c$. It is an important question whether with the introduction of a dynamical axion term the $p$-wave backgrounds remains thermodynamically preferred, however we will not study this here and focus on the interesting effects of the axion on the $p+ip$ transition with the hope to report on the above problem in future work.\\

Throughout this paper we decide to work in units of the $AdS$ radius, $L=1$ and the horizon radius $r_h =1$. Hence in this case $\frac{r}{L}, \frac{r}{r_h}\rightarrow r$ is a dimensionless parameter. Similarly, we can form dimensionless fields by  $\omega L\rightarrow \tilde{\omega}$ and $\phi L \rightarrow \tilde{\phi}$. Hence bear in mind that $\omega, \phi$ and $r$ are all dimensionful, as are all other fields in the theory, however the results presented in the paper are for dimensionless combinations of $\tilde{\omega}, \tilde{\phi}$ and $\frac{r}{r_h}$ with $r_h, L=1$ manually. \newline

The equations of motion one derives from the action are
\bea\label{eom}
\phi''+\frac{2}{r}\phi'-\frac{2}{r^4(1-\frac{1}{r^3})}\phi\omega^2+\frac{8}{r^2}\kappa\theta' \omega^2&=&0\\
\omega''+\frac{1+2r^3}{r(r^3-1)}\omega'-\frac{\omega^3}{r^4(1-\frac{1}{r^3})}+\frac{1}{r^4(1-\frac{1}{r^3})^2}\phi^2 \omega-\frac{8\kappa}{r^2(1-\frac{1}{r^3})}\theta'\phi\omega&=&0\\
\theta''+\frac{1+2r^3}{r(r^3-1)}\theta'+\frac{2}{r}\theta'-\frac{\partial_\theta V(\theta)}{2r^2\left(1-\frac{1}{r^3}\right)}-\frac{4\kappa}{r^4(1-\frac{1}{r^3})}(\phi\omega^2)'&=&0
\eea

where in the above $'$ denotes differentiation w.r.t $r$. Note that as per \cite{tallarita}, $\kappa\theta'$ acts as an effective Chern-Simons coupling in the boundary theory. The fields have asymptotic behaviour (again these should be thought as dimensionless fields at the horizon with $r$ replaced by $\frac{r}{r_h}$ everywhere)

\bea\label{horizon}
\tilde{\phi}&=&\tilde{\phi}_1(r-1)+\tilde{\phi}_2(r-1)^2+...\\
\tilde{\omega}&=&\tilde{\omega}_0+\tilde{\omega}_1(r-1)^2+... \nonumber\\
\theta&=&t_0+t_1 (r-1)+....\nonumber
\eea

at the horizon and 

\bea\label{boundary}
\tilde{\phi} &=& \tilde{p}_0 + \frac{\tilde{p}_1}{r}+...\\
\tilde{\omega}&=&\frac{\tilde{W}_1}{r}+....\nonumber\\
\theta&=&\theta_0+\frac{\theta_1}{r^3}+....\nonumber
\eea

at the asymptotic boundary at large $r$, where the expansion for $\theta$ is at $m=0$.  Through the gauge/gravity correspondence we can associate thermodynamic quantities to the above variables (\ref{boundary}) in the expansions of the fields at the asymptotic boundary. The holographic dictionary states that a bulk field is dual to a field theory operator on the boundary through the relation

\be
\bigg< exp \int \phi_0 \mathcal{O}\bigg> = exp (-S_{os}[\phi_0]),
\ee

where $S_{os}$ denotes the on-shell action. More precisely, the boundary value of the bulk field $\phi$ acts as a source for the corresponding field theory operator on the boundary. The method of calculation of the precise relationships between thermodynamic quantities of interest and bulk fields is illustrated fully in Appendix A of \cite{johnson}, to which we adhere. Therefore the charge density

\be
\rho=\frac{1}{2k_4^2g} r^2 \partial_r A_t \big|_{bry} = \frac{-1}{2k_4^2g} p_1,
\ee

and the chemical potential

\be
\mu=\frac{1}{2k_4^2g} A_t \big|_{bry} = \frac{1}{2k_4^2g}p_0.
\ee

note that here the asymptotic field values are untilded, hence they are dimensionful. In order to remove the dependence on $k_4$ from these quantities we make the redefinition $\hat{\rho}=\frac{k_4^2}{(2\pi)^3}\rho$ where the extra factors of $\pi$ are included for numerical convenience. Similarly the $AdS/CFT$ correspondence maps the Gibbons-Hawking  \cite{Gibbons} temperature $T_{GH}=\frac{3}{4\pi}$ of the black hole to the Temperature $T$ of the CFT and, as explained in detail in \cite{hartnoll3}, for a scale invariant theory at finite temperature in equilibrium there is no other meaningful scale which we can compare the temperature to, i.e. all non-zero temperatures should be equivalent. Hence, it is only physically meaningful to compare scale invariant quantities formed from the bulk fields projected on the boundary, to this purpose we define the effective temperature of the dual $CFT$ as  $\frac{T}{\sqrt{\hat{\rho}}}$, which from the above definitions is easily seen to be equal to

\be
\frac{T}{\sqrt{\hat{\rho}}}=3\sqrt{-\frac{\pi g}{2p_1}}.
\ee

As mentioned below equation (\ref{ansatz}), a non-vanishing value of $\omega$ at the boundary acts as a condensate for the dual field theory. Hence 

\be
J=-\frac{1}{2k_4^2g}r^2\partial_rA_x \big|_{bry} = \frac{1}{2k_4^2g}W_1 = \frac{1}{2g}\hat{W_1}
\ee

will be our order parameter, where in the last line we have made a similar redefiniton of $\omega$ to remove the dependence on $k_4^2$. As per the effective temperature defined previously we construct a physical scale invariant order parameter $\frac{J}{\hat{\rho}}$, which evaluates to

\be
\frac{J}{\hat{\rho}}=-\frac{\hat{W_1}}{p_1}.
\ee

In similar fashion, the bulk axion $\theta$ maps to a pseudo-scalar field theory operator on the boundary field theory, with $\theta_0$ acting as its source. The boundary behaviour of the axion field is such that
\be
\theta(r)= \theta_0 r^{-\Delta_{-}}+\theta_1r^{-\Delta_{+}}
\ee

where $\Delta_{\pm}$ indicate the dimensions of the corresponding boundary operators in the dual field theory. Note that as per a scalar field in $AdS$ space, the dimension of the boundary operator is dependent on $m$, the mass of the corresponding field in the bulk. In the case of $m=0$ then the asymptotics are those shown in \ref{boundary}, however for the case where $m\neq0$ then the analysis of varying $m$ shown below, which is for the case of $\theta_0$ not held fixed in the numerical procedure, corresponds in the field theory perspective as variations of the dual operator dimensions. \newline

To determine solutions to the equations of motion (\ref{eom}) we adopt a numerical ``shooting" procedure, which involves allowing for a non-vanishing constant term $W_0$ in the expansion for $\tilde{\omega}$ at the boundary and then manually ensuring that this vanishes\footnote{This is the normalisability condition for $\omega$.} by carefully changing the choices of $\tilde{\omega}_0$ and $\tilde{\phi}_1$. For the case of vanishing axion these three conditions are enough to determine a one parameter family of solutions which seeds the values at the asymptotic boundary, as explained in \cite{gubser2}. In the case of a non-vanishing axion we have three second order differential equations and thus six integration constants. The constraints are normalizability of $\omega$, the values of $\phi_1$ and $\omega_0$ and the values for $t_0$ and $t_1$, hence we have six constants and five constraints which makes a one-parameter family of solutions as per the previous case. We proceed in the numerical analysis working at constant $p_0$ (the grand canonical ensemble) and enforce that the asymptotic value of the axion $\theta_0$ is constant manually. For the case of a non-vanishing axion potential in which the asymptotics for $\theta$ cannot be controlled manually we proceed under the assumption that varying $m$ smoothly can enforce this condition, it is however not what has been done here and therefore this section should be taken only as a preliminary investigation on the effects the potential may have on forms of solutions with $\omega, \kappa \neq 0$. 

\section{Solutions}

This section is devoted to analysing numerical solutions of the above system. There are two variables which can be tuned by hand : $V(\theta)$ and $\kappa$ both subject to the overall solution being normalizable and thermodinamically preferred (see section below). We will start with the analysis for a vanishing axionic potential and proceed, in the next section, to include a mass term for the axion field.

\subsection{$V(\theta)=0$}

For the simplifying case of $V(\theta)=0$ we will observe the effect of raising the Chern-Simons parameter $\kappa$ on the space of possible solutions of the system, the profile for the axion field and the form of the condensate. The normalizable perturbations to the gauge field $\tilde{\omega}$ have many solutions with increasing nodes. We will restrict to solutions of the form shown in Figure 1 where $\tilde{\omega}$ is a monotonically decreasing function with no nodes as these are believed to be thermodynamically favoured over the other.\\

\begin{figure}[h]
\label{normpert}
\begin{center}
\includegraphics[scale=1]{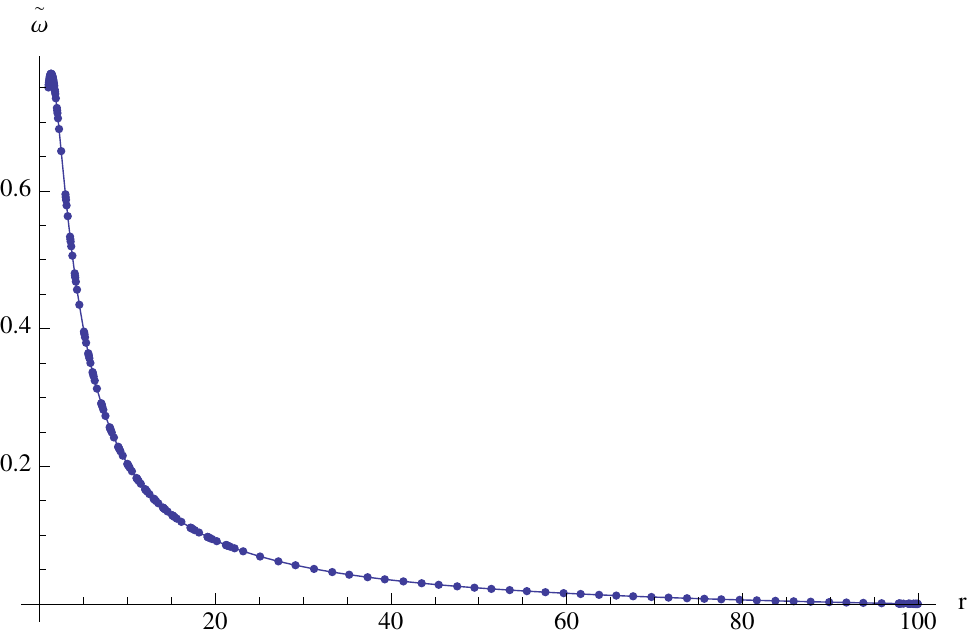}
\end{center}
\caption{Thermodynamically preferred form of the normalizable perturbation $\omega$ in which this is a monotonically decreasing function with no nodes.}
\end{figure}

Throughout this section we work with $t_0=1$ and $t_1=0$, we also drop the explicit tildes from the field values, bear in mind however that the quoted results are for dimensionless tilded forms of the fields. Figure 2 illustrates all possible values (to within the numerical accuracy of the procedure) for $\omega_0$ and $\phi_1$ which yield normalisable solutions with a non-vanishing condensate. The normal phase corresponds to $\omega_0=0$ whilst the phase with non-zero $\omega_0$ will be referred to as the superconducting phase. The blue/leftmost line corresponds to the case of $\kappa=0$, where the above ansatz for the gauge-field was shown to cause a second order phase transition between a RN black-hole and one with non-vanishing non-abelian condensate. Curves to the right of this are for increasing values of $\kappa$ in steps of 0.04. A maximum value of $\kappa=0.2$ is reached before the numerical procedure breaks down, at this value the CS term is large enough that one cannot ignore the back-reaction of the gauge fields. Note that in this scheme where $V(\theta)=0$ we see that all curves tend to the same value of $\phi_1$ as $\omega\rightarrow 0$. In this case one finds that $p_1$ is never greater than about 3.71, which leads to a constant ($\kappa$ independent) value of the critical temperature

\be\label{tc}
T_c \approx 1.95\sqrt{gL\rho}.
\ee

We will show in a later section that the independence of $T_c$ on $\kappa$ is justified analytically. 

\begin{figure}[h]
\label{m0vark}
\includegraphics[width=1\textwidth,height=0.35\textheight]{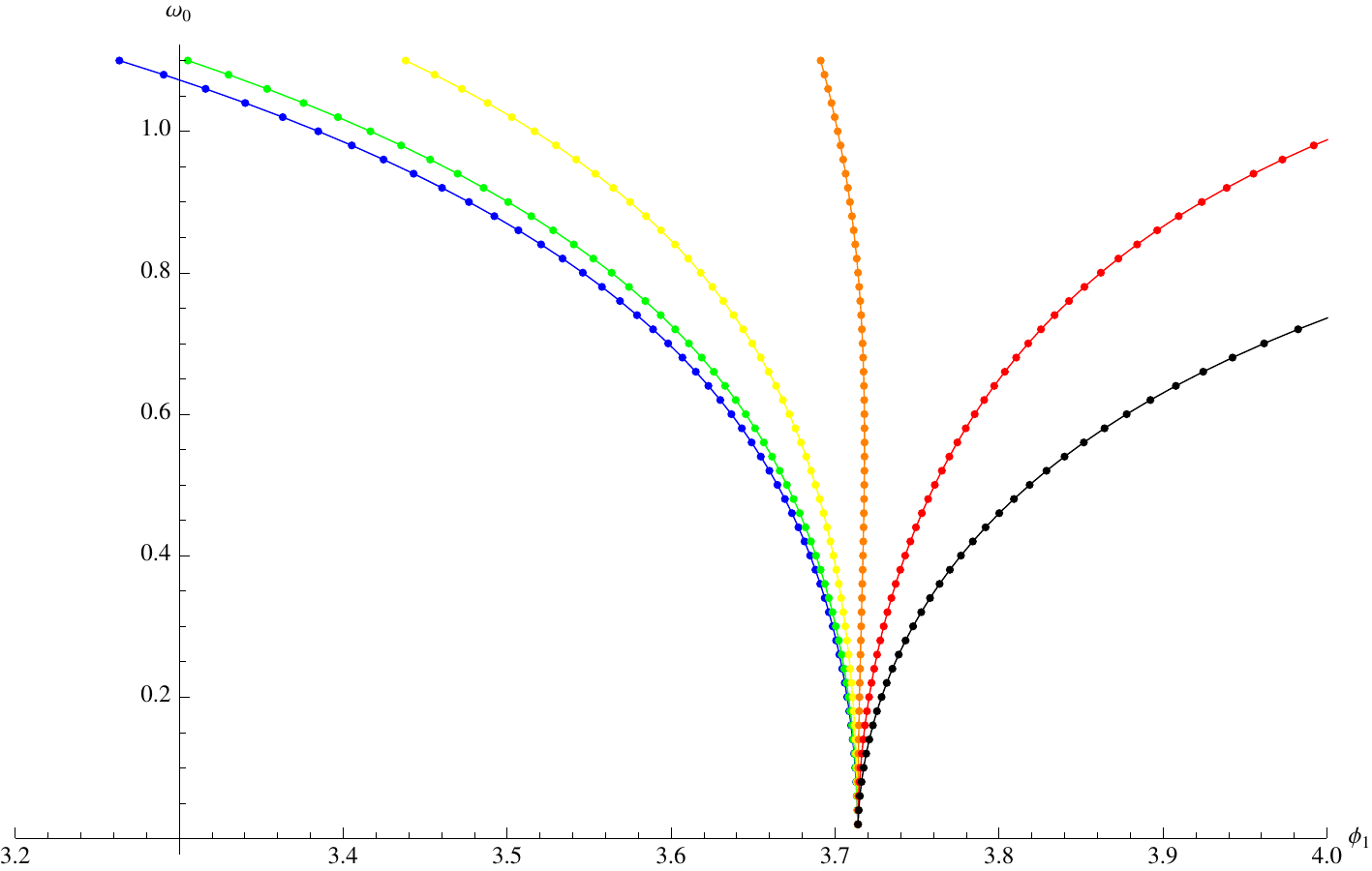}
\caption{The parameter space of possible solutions to the non-linear problem. The normal phase is described by $\omega=0$ whereas points on the line with non-zero $\omega$ describe the superconducting phase. As charge density and chemical potential are held constant, Temperature increases towards the left. The curve to the left is $\kappa=0$, then increasing $\kappa$ to the right in steps of 0.04. Hence the highest $\kappa$ shown in black is $\kappa = 0.2$.}
\end{figure}

\begin{figure}[h]
\label{normpert}
\begin{center}
\includegraphics[scale=1]{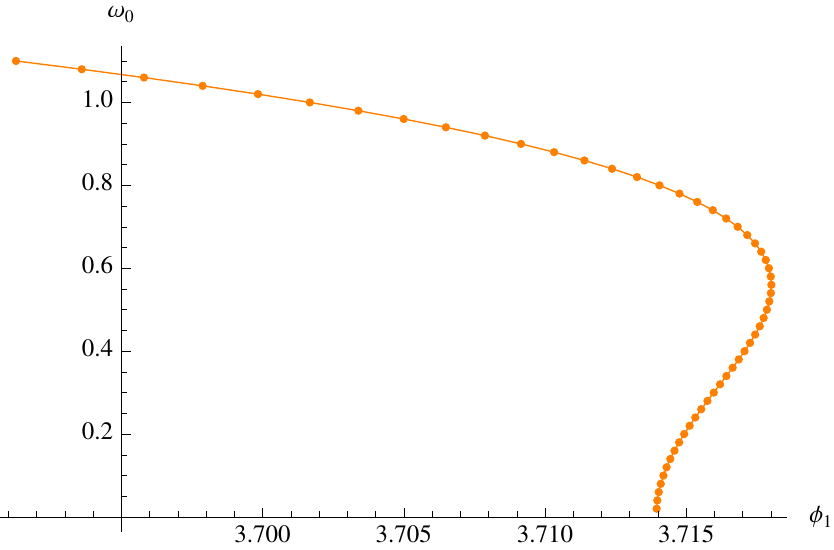}
\end{center}
\caption{First order phase transition at finite $\kappa=0.12$. The ``bump" seen extending beyond the region where the two phases meet at $\omega=0$ signifies that as one lowers the temperature there is a finite jump in the free energy between the two phases, and hence a first order phase transition.  }
\end{figure}

As is seen in Figure 2, for a narrow range of $\kappa$ the transition switches to first order. This is best seen in Figure 3 where the same plot is presented for $\kappa=0.12$, showing that the transition is first order, i.e. as one decreases the temperature for non-zero $\omega$ there is a non-continuous phase transition with a corresponding jump in the free energy.  Increasing $\kappa$ even further restores the transition to second order. The first order phase transition is seen for values of $\kappa$ ranging between $\kappa = 0.115$ to $0.123$.\\

The axion profile is shown in Figure 4. For $\kappa\neq0$, apart from a sharp rise around the position of the horizon we see that the axion is a constant everywhere. The different curves correspond to increasing values of $\kappa$, with the blue line corresponding to $\kappa=0$. The overall shape of the axion seems independent of $\kappa$. The fact that $\theta$ is a constant on the boundary does not mean that the black-hole develops axion-hair. One can effectively remove this by simply adding a constant term to the $F\wedge F$ term in the action. Solutions with the same asymptotic $\theta_0$ values for the axion field were compared here. Since in the case where $V(\theta)=0$ a constant axion profile is a solution to the equations of motion, the axion contribution disappears entirely from the normal phase, which is simply $AdS-RN$. In this case we are truly comparing a phase with no axion to that with an axion, this is not the case when we switch on a potential term as is discussed below. 

Finally, in Figure 5 we show the effect of different values of $\kappa$ on the condensate at the boundary. The plot shows the condensate as a function of $\frac{T}{T_c}$, with the highest line corresponding to the $\kappa=0$ case, then increasingly lower results at increasing $\kappa$ in steps of 0.04. The condensate suffers from a suppression as $\kappa$ increases, the form of which is investigated analytically in a later section.

\begin{figure}
\label{theta1}
\begin{center}
\includegraphics[scale=1.1]{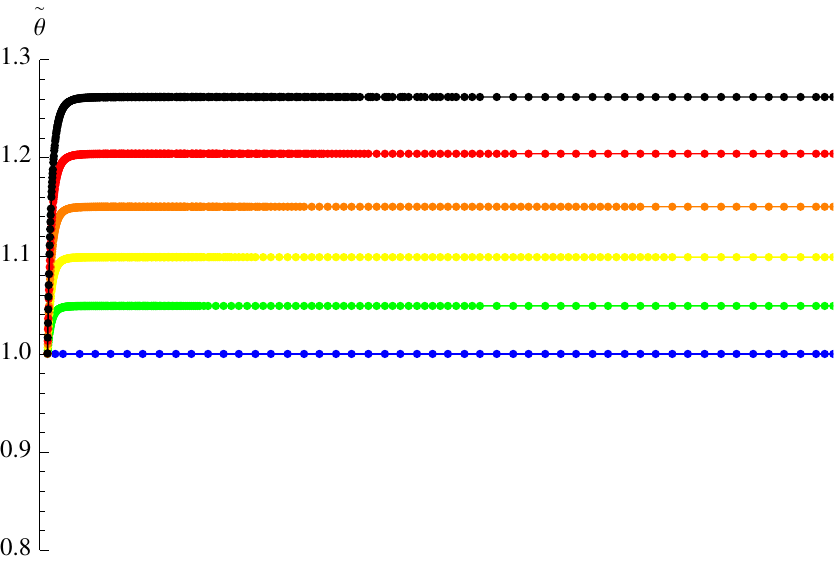}
\end{center}
\caption{The axion profile as a function of $\frac{r}{r_h}$, for a choice of $t_0 = 1$. The blue/lowest line is $\kappa=0$, then increasing $\kappa$ upwards in steps of 0.04.}
\end{figure}

\begin{figure}[h]
\label{cond}
\begin{center}
\includegraphics[scale=1]{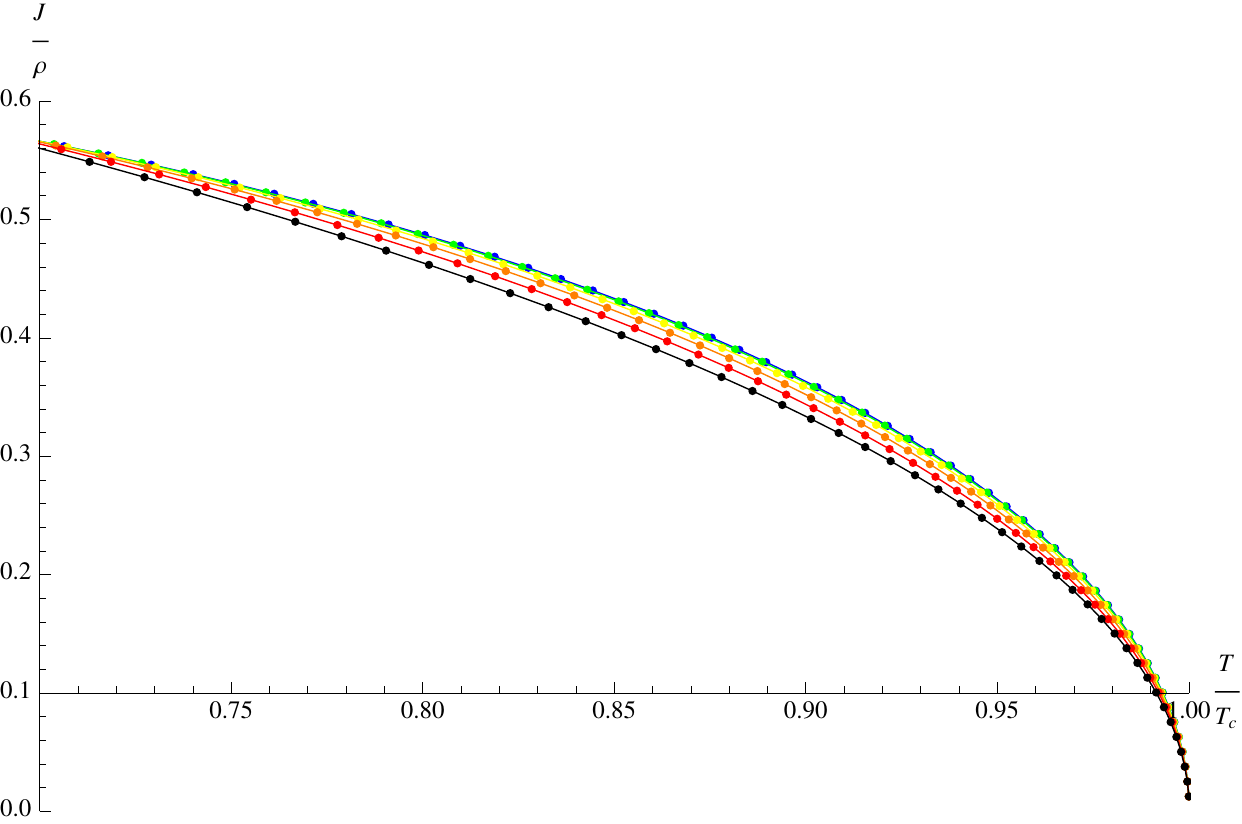}
\end{center}
\caption{The condensate profile for increasing $\kappa$. The highest line corresponds to $\kappa=0$, then increasing $\kappa$ downwards in steps of 0.04. There is a small suppression in the condensate profile for increasing $\kappa$.}
\end{figure}

\subsection{$V(\theta)\neq0$}

In this section we switch on a mass for the axion field so that $V(\theta)=m^2\lambda^2\theta^2\rightarrow m^2\theta^2$ and we will investigate what happens to the phase transition as one varies the mass $m$ at finite non-zero small $\kappa$, we restrict the analysis to $\kappa =0.08$, below the region where we see a first order phase transition. Note that with the contribution from a potential term a constant axion profile is not a solution to the equations of motion (\ref{eom}), as is indeed seen in Figure 7 even for the simple case of $\kappa=0$. This implies that we have a non-zero contribution from the axion kinetic and potential terms present in the ``normal" phase where $\omega=0$. The comparison is thus between an $AdS$-Schwarzschild black hole with a non-trivial profile for the axion to a ``superconducting" phase with a condensate and an axion. Whilst the condition of comparison of the same asymptotics for the $\theta$ field is easily met in the case of vanishing potential, this cannot be done straightforwardly here. In this case the asymptotic value of the axion field varies as I vary the mass $m$, this is because the profile of the axion varies as a function of $m$ and therefore the extracted values of $\theta_0$ will be different for each profile of the axion in the case where this is not held constant in the numerical procedure, which applies here. Hence comparison with the same asymptotic for both phases requires smooth variations of $m$ with $\omega$ so as to maintain $\theta_0$ constant. This has not been done in this section, which rather illustrates how varying $m$ values changes the form of solutions with differing values of $\theta_0$ with constant $\omega$ and constant $\kappa$. It remains an important question which solution is thermodynamically preferred when the axion has a potential term contribution, whether a phase transition occurs in the first place and whether this term has noticeable effects on the space of solutions of the system. These are the issues which this preliminary investigation wishes to address.\newline

In Figure 6 we show the effect of raising $m$ on the space of possible solutions of the system. The blue/leftmost curve corresponds to the case of vanishing $m=0$ and curves to the right of this are for increasing values of $m$ in steps of 0.1, the final curve is $m=0.45$ as already for $m=0.5$ the back-reaction cannot be ignored and the numerical procedure breaks down. We see that increasing the mass of the axion has the effect of shifting the phase transition curve to higher values of $\phi_1$ as $w\rightarrow0$ whilst preserving its shape. This has the interesting effect of lowering the critical temperature $T_c$ at which the transition takes place. Unfortunately we are restricted from investigating the region of large $m$ from the numerical procedure. With variations in $m$ we also observe a variation in the shape of the axion profile. This is shown in Figure 7 where, contrary to changing $\kappa$, the axion has a non-trivial profile in the bulk and changing $m$ doesn't correspond to a simple shift for $\theta$.\newline

\begin{figure}[h]
\label{varyingm}
\begin{center}
\includegraphics[scale=1.2]{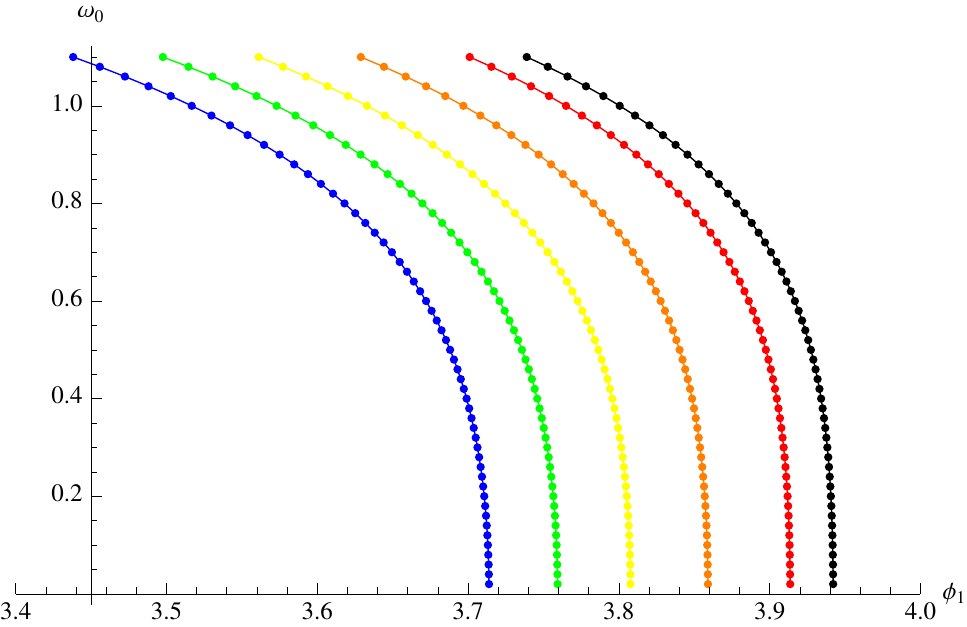}
\end{center}
\caption{The effect of the mass $m$ on the space of possible solutions at $\kappa=0.08$. The blue/leftmost curve corresponding to $m=0$ then increasing $m$ towards the right in steps of 0.1, the final curve is $m=0.45$.}
\end{figure}

It is interesting to observe the effects of $m$ on the order parameter of the field theory. We saw in the previous section that when $m=0$, varying $\kappa$ had (to within the tested numerical range of parameters) a small effect on the $T\approxeq T_c$ region of the order parameter. In Figure 8 we have plotted the order parameter against $\frac{T}{T_c}$ for curves with different values of $m$. The blue/highest curve is the $m=0$ case and the curves below this are for increasing $m$. Each curve is plotted with its corresponding value of $T_c$. These results are all for the choice $t_0=1$. Given that $t_0$ appears coupled to $m$ as the Chern-Simons interaction term on the boundary it is evident that fixing $m\neq 0$ means that varying $t_0$ has analogous effects to the system to varying $m$.

\begin{figure}[h]
\label{varmtheta}
\begin{center}
\includegraphics[scale=1]{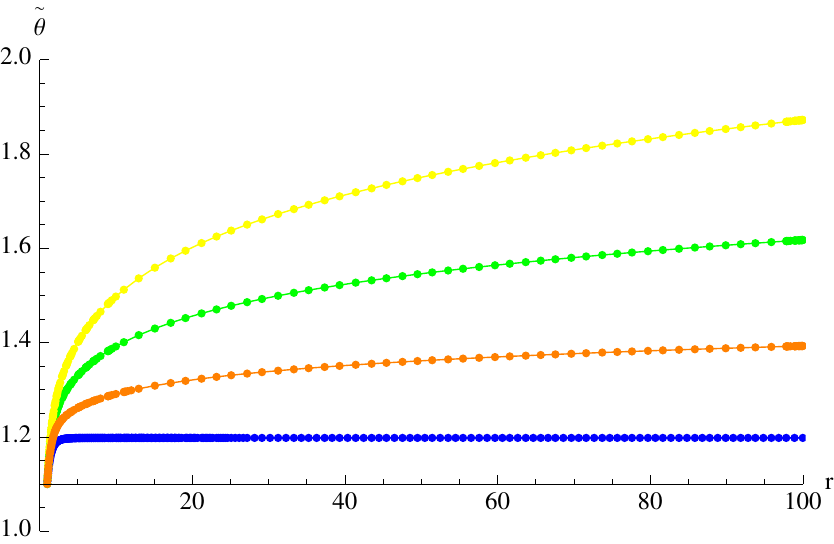}
\end{center}
\caption{The effect of the mass $m$ for the axion field at $\kappa=0.08$. Increasing $m$ upwards in steps of 0.1, the blue/lowest curve corresponds to $m=0$. }
\end{figure}

\begin{figure}[h]
\label{condvarm}
\includegraphics[width=0.9\textwidth,height=0.35\textheight]{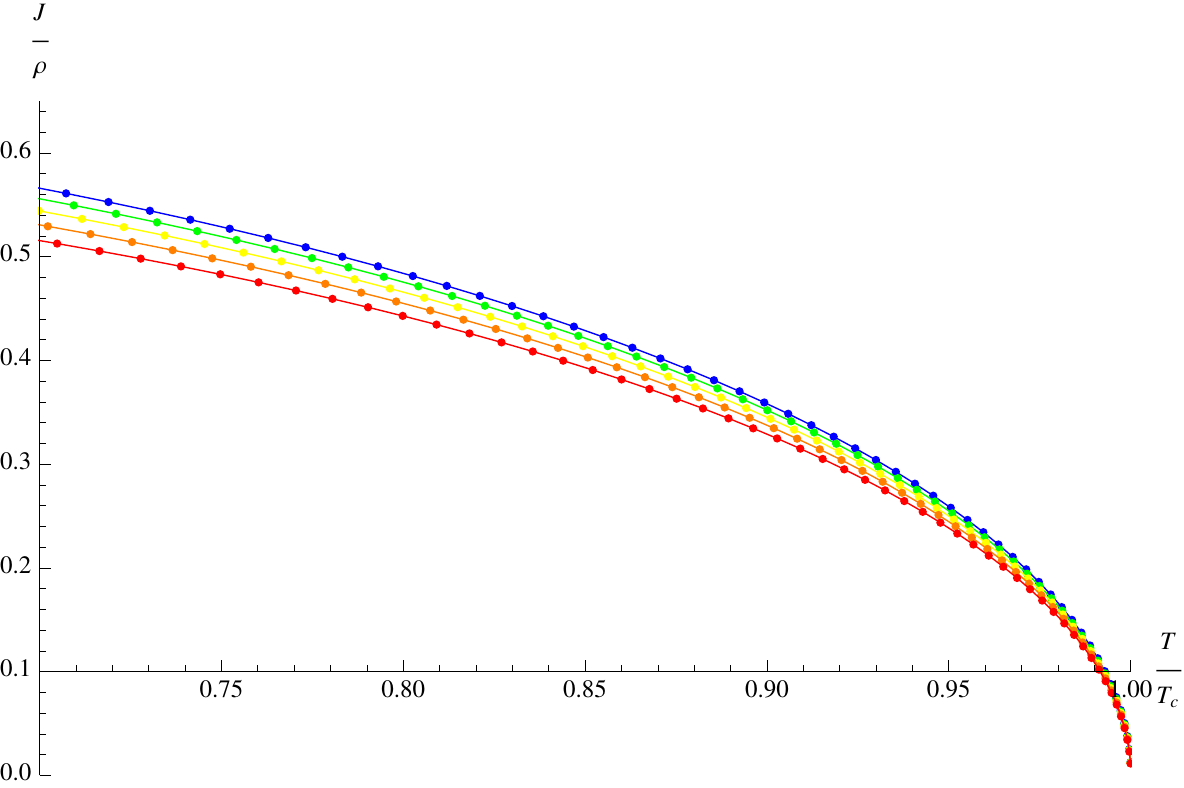}
\caption{Increasing $m$ effects on condensate at fixed $\kappa=0.08$. The highest/blue curve corresponds to $m=0$, the lower curves represent increasing $m$ in steps of 0.1. There is a significant effect on the profile of the condensate.}
\end{figure}

\section{Free Energy}

In this section we make use of the $AdS/CFT$ correspondence to calculate the free energy density for the superconducting phase $f$. We are interested in the scale invariant quantity $\frac{\Delta f}{\rho^{1.5}} = \frac{f-f_{RN}}{\rho^{1.5}}$, the difference in free energy densities between the normal and the superconducting phase. If this remains negative, then the symmetry breaking phase is preferred.\\

By the $AdS/CFT$ dictionary the free energy is given by the on-shell Euclidean action with appropriate counter terms to cure divergences. Therefore we work with
\bea
2k_4^2 S &=& S_{grav}+S_{Maxw}+S_{CS}
\eea

where 

\be
S_{grav}=\int d^4x \sqrt{-g}\left(R+6\right)+\int_{r_{\infty}} d^3x\sqrt{\gamma}\left(-2K+4\right),
\ee

the second term is the usual Gibbons-Hawking \cite{gibbons2} boundary term needed to have a sensible variational principle, $\gamma$ is the induced metric on the boundary and we also add a boundary cosmological constant term to regulate the action, then 

\be
S_{Maxw}=-\int d^4x\frac{\sqrt{-g}}{4}Tr\left(F_{\mu\nu}F^{\mu\nu}\right)
\ee

is the usual Maxwell term with no further counter-terms needed as we assume that the gauge field goes to zero at the boundary sufficiently quickly, finally

\be
S_{CS}=\int d^4x \sqrt{-g}\left(\partial_\mu\theta\partial^\mu\theta+V(\theta)\right)+\int d^4x \kappa\theta Tr\left(F\wedge F\right)
\ee

where these terms are the usual kinetic, potential and Chern-Simons terms for the axion field. We then proceed to evaluate this action on-shell with the Euclideanised $t\rightarrow i\tau$ metric to obtain

\bea\label{free}
\Delta f&=&\left(-\frac{p_0^2}{4}+f_{CS}^0\right)+\int_{r_{h}}^{\infty} dr\Big[\frac{1}{4}r^2(\phi')^2-\frac{1}{2}r^2\left(1-\frac{r_h^3}{r^3}\right)(\omega')^2-\frac{1}{4r^2}\omega^4\Big.\nonumber\\
&&\Big.-\frac{1}{2r^2\left(1-\frac{r_h^3}{r^3}\right)}(\phi\omega)^2+8i\kappa\theta(\omega^2\phi)'+ r^4\left(1-\frac{r_h^3}{r^3}\right)(\theta')^2+r^2V(\theta)\Big]
\eea

where the first term is given by the normal phase Maxwell term, $f_{CS}^0$ denotes the CS terms evaluated in the normal phase and the remaining terms are obtained from the numerical solutions. The terms in $f_{CS}^0$ are simply the kinetic and potential axion terms in the normal phase, whose presence or disappearance is determined non-trivially as they don't have direct couplings to $\omega$  (which vanishes in this phase).

\subsection{$V(\theta)=0$}

 From the equations of motion (\ref{eom}) we can see that in the case of vanishing axion mass $m=0$ then the kinetic term in the normal phase, which is proportional to $\theta'$ doesn't contribute as a consistent solution for the axion is simply to have a constant profile. Hence in this case, $f_{CS}^0 =0$ and we don't have to worry about CS contributions to the free energy coming from the normal phase. We now proceed to evaluate the remaining terms in order to obtain the overall change in free energy. Let's start with the term involving $\kappa$, this is 
\be
f_k = \int_{r_h}^{\infty} dr 8 i\kappa \theta (\omega^2\phi)'
\ee

and hence when we Euclideanise the action, this term acquires a factor of $i$ from the $dt$ component of the gauge field. This means that the term is irrelevant in the partition function $e^{-S}$ and thus doesn't contribute to the free energy. This remains true in the case of a non-zero axion potential.\\

The kinetic term for the axion field contributes to the free energy density in the form

\be
f_{\theta}=\int_{r_h}^{\infty} dr r^4\left(1-\frac{r_h^3}{r^3}\right)(\theta')^2
\ee

which contributes a finite amount to the free energy, changing with $\kappa$ as the corresponding solution for $\theta$ changes. It might at first sight appear surprising that this term remains finite over the integration due to the direct coupling to a factor of $r^4$ which diverges at infinity, however we can see from the asymptotic behaviour of the axion \ref{boundary} $(\theta')^2 \approx r^{-8}$ that this factor is cancelled in the overall integral, which thus remains finite. A sketch of the profile of the derivative of the axion term is shown in Figure 9, note how the axion decays to zero for large $r$. Figure 10 shows the effects that increasing $\kappa$ has on the free energy through the changes in profile of the axion term. For the values of $\kappa$ tested in the numerical procedure, the curve remains negative and thus the axion phase is preferred.

\begin{figure}[h]
\label{condvarm}
\begin{center}
\includegraphics[scale=0.7]{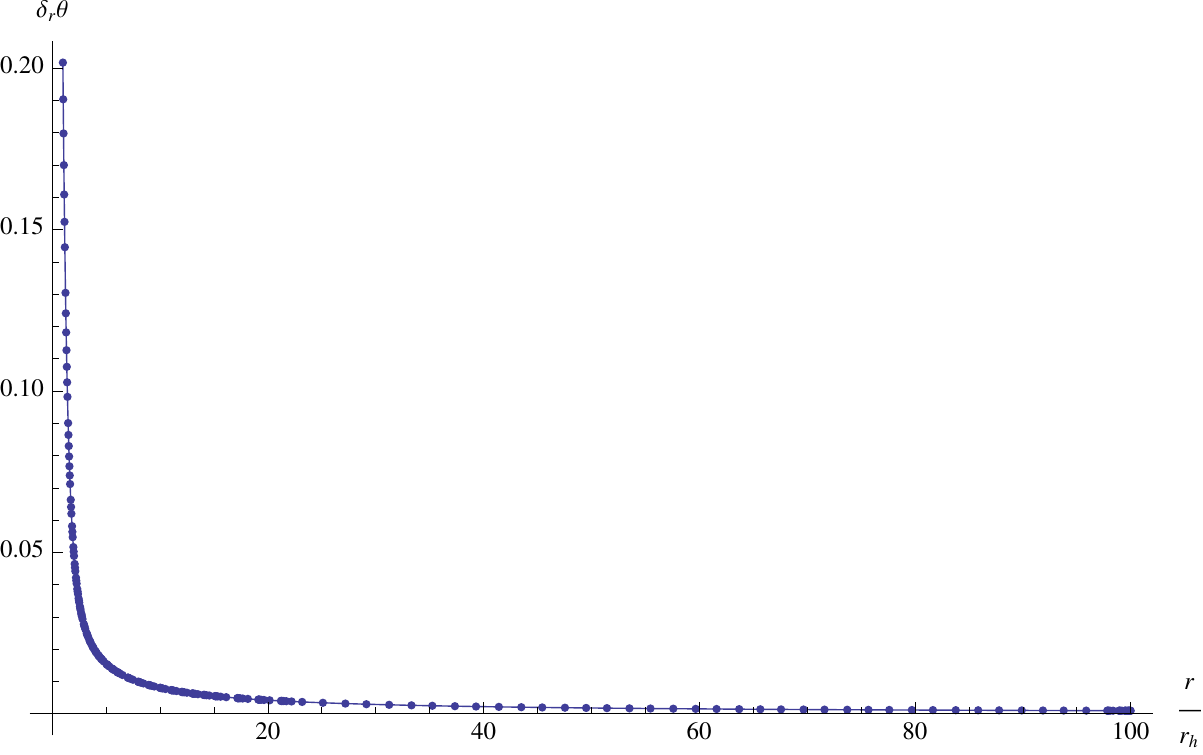}
\end{center}
\caption{The numerical solution for $\theta'$. At large $r$ the profile is vanishing, this illustrates how the kinetic contribution of the axion term remains well behaved despite having explicit couplings to positive powers of $r$.}
\end{figure}

\begin{figure}[h]
\label{freeenfork}
\begin{center}
\includegraphics[scale=0.7]{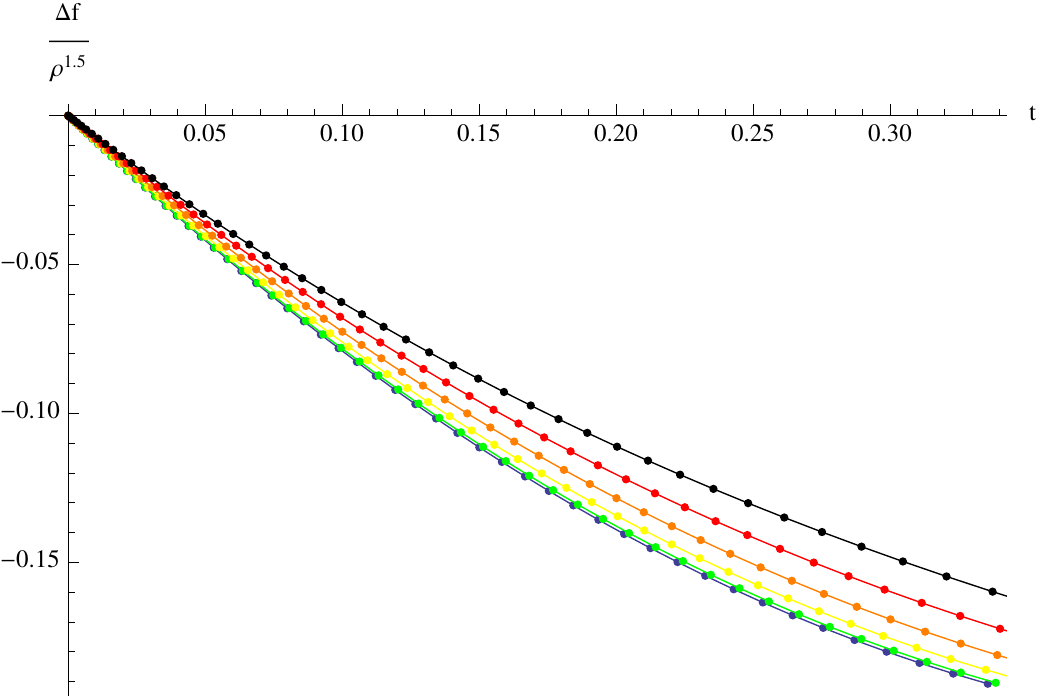}
\end{center}
\caption{The effect of increasing $\kappa$ on the free energy density. The lowest blue curve corresponds to $\kappa=0$ then increasing $\kappa$ upwards in steps of 0.04. For all $\kappa$, $\Delta f$ remains negative.}
\end{figure}

\subsection{$V(\theta)\neq0$}

In the case of non-vanishing potential things become greatly more complicated. CS contributions are present in the normal phase as $\theta$ having a constant profile is no longer a solution to the equations of motion. Hence in this case the normal phase contributes 
\be
f_{No} = -\frac{p_0^2}{4}+ \int_{r_h}^{\infty} dr \left(r^4\left(1-\frac{1}{r^3}\right)(\theta')^2+ r^2 V(\theta)\right)\big|_{\omega=0}
\ee

and we have to make sure we consider the resulting kinetic contribution arising between the difference in kinetic axion terms in both the normal and superconducting phases. Most importantly the term
\be
f_{V(\theta)}=\int_{r_h}^{\infty} dr r^2 V(\theta)
\ee

is divergent as $V(\theta)=m^2\theta^2$ and $\theta$ is non-vanishing at asymptotic $r$. One can hope that the divergent contribution from the normal phase here can cancel that of the superconducting phase, however this cannot be the case exactly since the two phases have different profiles for the axion $\theta$. Therefore to make progress we could proceed in two distinct ways: by a systematic holographic renormalization procedure (see \cite{kostas}) in which the necessary counterterms are computed manually and the divergences are cancelled explicitly or by ensuring that the numerical procedure is such as to maintain a constant value for $\theta_0$ at the boundary, in which case the divergences should cancel exactly in both phases (clearly both methods should eventually agree). This section is intended as an introduction to the case of $V(\theta)\neq0$ and thus we won't carry out the manual renormalization procedure here.  Therefore, it remains an open and important questions whether a non-zero mass $m$ for the axion can de-stabilise the ``superconducting" phase or not. We wish to report on this (and especially on the second numerical method for cancelling the divergences) in future work.

\section{Analytic Calculations}

Even though the set of equations (\ref{eom}) require numerical solutions, one can still obtain an analytical feel for the behaviour of thermodynamic quantities of interest by looking at the region close the phase transition. In this region the fields are small so one can trust the series solutions for the fields obtained by matching the series expansions at the horizon and at the boundary, the procedure is detailed in \cite{herzog3}. Even though it is an excellent method to gain analytic understanding for the behaviour of quantities of interest this procedure is only an approximation valid in the region of small fields, hence one should be sceptical about the precise numerical results presented. By inserting equations (\ref{horizon}) into (\ref{eom}) and matching to the expansions at the boundary we obtain for the case of $m=0$ that
\bea
T_c=C\sqrt{ gL\rho},
\eea

where $C$ is an arbitrary proportionality constant chosen in the analytical procedure. By choosing $C=1.95$ we can recover the numerical result \ref{tc}.  So the critical temperature doesn't depend on $\kappa$ as is observed in Figure 2, and

\be
\frac{J}{\rho}=\frac{2(gL)^2\sqrt{\left(1-\frac{T}{T_c}\right)}}{\sqrt{3}\sqrt{\frac{7}{2}+96\kappa^2}}\\
\ee

which shows the dependence of the condensate on $\kappa$. The precise numbers appearing in this relations are unimportant, it is intended to demonstrate the explicit behaviour with varying $\kappa$. It also indicates that the transition at finite $\kappa$ has simple critical exponents given that $T_c$ is $\kappa$ independent. This should be trusted close to the transition in the vicinity of $T=T_c$. From Figure 5 we can see that this analytical behaviour matches well with the numerical results. Switching on a mass term for the axion gives a complicated expression for $T_c(m)$ which we won't include for simplicity, however it is important to see that even analytically when we include a non vanishing potential for the axion field then the critical temperature shifts as a function of the mass, as observed in Figure 6. Furthermore, this indicates that in this case the transition may not have the simple $\sqrt{1-\frac{T}{T_c}}$ behaviour close to the transitions, depending on the value of $m$.\\

We must also impose that the boundary Chern-Simons term is quantized. The requirement 
\be
\kappa \theta_{r_{\infty}} = \frac{n}{4\pi}
\ee

is translated to a requirement on the series expansion at the horizon for $\theta$ by the analytical procedure. We obtain that

\be
t_0 = \frac{n}{16\pi\kappa}-\frac{96 t\kappa}{7+192\kappa^2},
\ee

this makes sense in this approximation where $T$ is close to $T_c$ and $t\approxeq0$. Working with $t_0=1$ in the numerical procedure amounts to a rescaling of the analytical $t_0$ by $16\pi\kappa$ and working at $n=1$.

\section{Conclusions}

In this paper we have investigated the effects of adding a Chern-Simons term, coupled to an axion field, to the phase transition between $AdS$ Reissner-Nordstrom and $AdS$ black-holes with a non-abelian condensate. In particular we considered the two cases of vanishing potential for the axion field and providing this with a finite mass $V(\theta)=m^2\lambda^2\theta^2$. In the first case we showed that increasing $\kappa$ led to a transition which alternates between first order and second order. As $\omega\rightarrow0$ all solutions ended at equal $\phi_1$ and no change in $T_c$ was observed. In terms of the order parameter we showed that this was suppressed with increasing $\kappa$, even though for the range of $\kappa$ tested the effects were small. We showed analytically that the condensate has a $(a+b\kappa^2)^{-\frac{1}{2}}$ for constant $a,b$ dependence which matches well with our numerical results. There were no noticeable changes in the results when varying the choice for the expansion of $\theta$ at the horizon. With calculations of the free energy we showed that the superconducting phase remains dominant below $T_c$.\\
\indent When the axion field is given a finite mass things are significantly different. Firstly, the axion field develops a non-constant profile in the bulk which makes the asymptotic comparison of the two phases difficult, the space of solutions which admit a non-vanishing $\omega$ is shifted to higher values of $\phi_1$ as $\omega\rightarrow0$. This in turn has the effect of lowering the critical temperature $T_c$ at which the transition takes place. Being non-constant the axion contributes to the normal phase as well, hence in this case the transition is between an $AdS/RN$ black hole with some axion hair to a ``superconducting" phase with axion hair as well. Also, as pointed out inside the paper, the asymptotic matching of the axion is hard when the potential is present and therefore the results shown are only for constant $\omega$ and $\kappa$ but varying $\theta_0$. Indeed it is crucial to extend the algorithm used for the results of this paper to one which can maintain constant $\theta$ values at the boundary. Even though this might appear a straight-forward task conceptually (after all it is the addition of one further constraint to the numerical solver) the implementation in terms of the numerical procedure presents problems due to the highly non-linear nature of the axion interaction terms. To make a true comparison between phases in which $\omega=0$ to those called ``superconducting" one would have to maintain the same boundary asymptotics for $\theta$ which requires smooth variations of $m$ as $\omega$ is varied. This is an important task which certainly should be investigated further. Nevertheless, the presence of the potential has significant effects on the characteristics of the phase transition at constant $\omega$, suppressing the order parameter. Similarly, one would like to show that for all regions of $m$ the free energy contribution of this potential term is small enough so as to not make the overall energy change positive, signifying that the transition is not preferred. Once the constant asymptotic values for the axion are maintained the divergences should cancel in both ``superconducting" phases and the results should be reliable. We wish to report on this area of work in future work.\\
\indent	In conclusion we have observed that a CS term has important effects on the phase transition of $AdS$ black-holes with non-abelian condensates. It would be desirable to investigate the region of large $m$ and large $\kappa$, where in the most optimistic case (with high-precision numerics) one could also push $T_c$ down to zero. As is apparent from the paper, the most important obstacle in making progress with this system is the complexity of the numerical procedure. One would really like to be able to probe the limiting regions of the parameter space. For example we would like to investigate the interesting region of large $\kappa$ where the system is indicative of admitting a region of non-vanishing $\omega$ for large $\phi_1$. The numerical procedure is reliable up to the point when terms become large enough to back-react significantly on the geometry and the approximation breaks down.\\
\indent	One very important direction for further research involves modifications of the gauge field ansatz. The next step is to investigate the effects of the CS term in the context of p-wave Holographic Superconductivity \cite{gubser2}, this requires a less istropic form for the gauge field. One way to make progress is to include constant values in both directions of the gauge field ansatz, i.e. $A \approx \omega(a\tau^1dx+b\tau^2dy)$ for $a,b$ constants and vary these numerically. Then one can hope that in regions where $a>>b$ we could drive the system to resemble one which has fewer isotropic characteristics. \\
\indent	Regarding the above system, one immediate desire is to push the analysis to the fully back-reacting case of finite $g$ in which one would observe the complete effect of the CS term. Furthermore, one can freely change the profile for the axion potential $V(\theta)$, for example one could include a $\tilde{\lambda}\theta^4$ term, this may lead to novel interesting features not observed here. In general, the axion potential system deserves detailed analysis with regards to the free energy and the asymptotics, where all scaling symmetries of the equations of motion could determine whether a comparison between the two phases is reliable even without explicit matching of $\theta_0$. The $\theta F\wedge F$ term can arise, for example, in anomaly cancellation in string-theory. It remains an open and interesting problem to find a suitable string-theory reduction which leads to the above set-up.

\section{Acknowledgments}

The author would like to thank J.Pasukonis for technical help, D.Rodriguez-Gomez for patient explanations and F. Dell'Aria for support. The work of G.T is supported by an EPSRC studentship.

\end{document}